\documentclass[10pt,journal,compsoc]{IEEEtran}
\usepackage{amsmath,amssymb,amsfonts}
\usepackage{algorithmic}
\usepackage{graphicx}
\usepackage{textcomp}
\usepackage{xcolor}
\def\BibTeX{{\rm B\kern-.05em{\sc i\kern-.025em b}\kern-.08em
    T\kern-.1667em\lower.7ex\hbox{E}\kern-.125emX}}

\usepackage{array, boldline, makecell, booktabs} 
\usepackage{tikz}    

\usepackage{todonotes}
\usepackage{arydshln}
\usepackage{adjustbox}

\usepackage{pgfplots}
\usepackage{subfigure}
\usetikzlibrary{patterns}
\usetikzlibrary{pgfplots.groupplots}
\usetikzlibrary{decorations.pathmorphing}
\usepackage{arydshln} 
\usepackage{float}
 
\usepackage{url}
\usepackage{amsmath,amsfonts,latexsym,amssymb,xspace}

\usepackage{relsize}
\usepackage{times}
\usepackage{cases}
\usepackage{paralist}
\usepackage{subfigure}
\usepackage[mathscr]{eucal}

\usepackage{stmaryrd}
\usepackage{bm}
\usepackage{textcomp}
\usepackage{epsfig}

\usepackage{arydshln} 
\usepackage{epstopdf}

\usepackage{stfloats}
\usepackage{tikz}
\usetikzlibrary{shapes}
\usetikzlibrary{matrix}
\usetikzlibrary{patterns}
\usetikzlibrary{backgrounds}
\usepackage{pgfplots}
\pgfplotsset{compat=1.3}
\usetikzlibrary{scopes}

\usepackage{sansmath}

\newcommand{\bval}[1]{$\mathsf{bval}(#1)$\xspace}

\newcommand{\upon}{{\vspace{0.1pt}{\textbf{upon receiving}}}\xspace}

\newcommand{\figref}[1]{Figure~\ref{#1}\xspace}

\newcommand{\mib}{MiB\xspace}
\newcommand{\miba}{MiB5\xspace}
\newcommand{\mibb}{MiB7\xspace}
\newcommand{\mbc}{MBC\xspace}
\newcommand{\mbcl}{MBC-L\xspace}
\newcommand{\mavid}{AVID\xspace}

\newcommand{\avid}{AVID\xspace}
\newcommand{\avidl}{AVID-L\xspace}

\newcommand{\mibaa}{MiB5a\xspace}
\newcommand{\mibab}{MiB5b\xspace}
\newcommand{\mibba}{MiB7a\xspace}
\newcommand{\mibbb}{MiB7b\xspace}
\newcommand{\mibbc}{MiB7c\xspace}

\usepackage{amsmath}
\newtheorem{theorem}{Theorem}

\newcommand{\IF}{{\vspace{0.1pt}{\textbf{if}}}\xspace}

\newenvironment{packeditemize}{
\begin{list}{$\bullet$}{
\setlength{\labelwidth}{8pt}
\setlength{\itemsep}{0pt}
\setlength{\leftmargin}{\labelwidth}
\addtolength{\leftmargin}{\labelsep}
\setlength{\parindent}{0pt}
\setlength{\listparindent}{\parindent}
\setlength{\parsep}{0pt}
\setlength{\topsep}{3pt}}}{\end{list}}

\newcommand{\heading}[1]{{\vspace{3pt}\noindent{\textbf{#1}}}}
\newcommand{\headit}[1]{{\vspace{1pt}\noindent{\textit{#1}}}}

\newcommand{\noskipheading}[1]{{\noindent{\textbf{#1}}}}

\newcounter{stepcnt}
\ifCLASSOPTIONcompsoc
  \usepackage[nocompress]{cite}
\else
  \usepackage{cite}
\fi

\ifCLASSINFOpdf
 
\else
  
\fi

\begin{document}

\title{MiB: Asynchronous BFT with More Replicas}

\author{Chao~Liu,
        Sisi~Duan,
        and~Haibin~Zhang
\IEEEcompsocitemizethanks{
\IEEEcompsocthanksitem C. Liu is with the Department
of Computer Science and Electrical Engineering, University of Maryland, Baltimore County, Baltimore,
MD, 21250. 
E-mail: chaoliu717@umbc.edu
\IEEEcompsocthanksitem S. Duan is with Tsinghua University. Corresponding author.
E-mail: duansisi@mail.tsinghua.edu.cn
\IEEEcompsocthanksitem H. Zhang is with Shandong Institute of Blockchain. Corresponding author.
E-mail: bchainzhang@aliyun.com}
}

\IEEEtitleabstractindextext{%
\begin{abstract}
State-of-the-art asynchronous Byzantine fault-tolerant (BFT) protocols, such as HoneyBadgerBFT, BEAT, and Dumbo, have shown a performance comparable to partially synchronous BFT protocols. This paper studies two practical directions in asynchronous BFT. First, while all these asynchronous BFT protocols assume optimal resilience with $3f+1$ replicas (where $f$ is an upper bound on the number of Byzantine replicas), it is interesting to ask whether more efficient protocols are possible if relaxing the resilience level. Second, these recent BFT protocols evaluate their performance under failure-free scenarios. It is unclear if these protocols indeed perform well during failures and attacks.

This work first studies asynchronous BFT with suboptimal resilience using $5f+1$ and $7f+1$ replicas. We present \mib, a novel and efficient asynchronous BFT framework using new distributed system constructions as building blocks. \mib consists of two main BFT instances and five other variants. As another contribution, we systematically design experiments for asynchronous BFT protocols with failures and evaluate their performance in various failure scenarios. We report interesting findings, showing asynchronous BFT indeed performs consistently well during various failure scenarios. In particular, via a five-continent deployment on Amazon EC2 using 140 replicas, we show the \mib instances have lower latency and much higher throughput than their asynchronous BFT counterparts.
\end{abstract}

}

\maketitle

\IEEEdisplaynontitleabstractindextext

\IEEEpeerreviewmaketitle

\section{Introduction}\label{introduction}

State machine replication (SMR) is a popular software technique achieving high availability and strong consistency guarantees in today’s distributed applications (e.g., Apache ZooKeeper~\cite{hunt2010zookeeper}), Google's Spanner~\cite{burrows2006chubby}). 
Byzantine fault-tolerant SMR (BFT) is known as \textit{the} model for permissoned blockchains, where the replicas 
need to authenticate themselves but
do not necessarily trust each other. 
BFT is also used in permissionless blockchains (where nodes may join and leave the systems dynamically) to enhance the performance and achieve finality (e.g.,~\cite{byzcoin,pass2016hybrid,rapidchain,elastico,omniledger,solida}). 

Different from partially synchronous BFT protocols, asynchronous BFT protocols do not rely on any timing assumptions and are therefore more robust against performance, timing, and denial-of-service attacks. For this reason, many asynchronous BFT (atomic broadcast) protocols have been proposed~\cite{ben1983another,cachin2001secure,ben1994asynchronous,ritas2008,sintra,ks,rc,abraham2018validated,correia2006consensus}. 
In particular, several asynchronous BFT protocols proposed recently, HoneyBadgerBFT~\cite{miller2016honey}, BEAT~\cite{duan2018beat}, EPIC~\cite{EPIC2020}, 
and Dumbo~\cite{Dumbo2020}, have comparable
performance as partially synchronous BFT protocols (e.g., PBFT~\cite{castro2002practical}) and can scale to around 100 replicas.   
These efficient protocols follow the asynchronous common subset (ACS) framework~\cite{ben1983another}. The ACS framework consists of a reliable broadcast (RBC) component and an asynchronous binary agreement (ABA) component. 
HoneyBadgerBFT, BEAT, EPIC, and Dumbo all use RBC and ABA and are different only in concrete instantiations. 

All these efficient asynchronous BFT protocols using RBC and ABA assume \textit{optimal resilience}. Namely, if the system has $n$ replicas, it tolerates $f < n/3$ Byzantine failures. Even in the best-case scenario, each asynchronous BFT epoch has at least six steps and the expected number of steps is much higher.\footnote{In the best-case scenario, ACS includes a RBC phase with $n$ (parallel) RBC instances and an ABA phase with $n$ (parallel) ABA instances. RBC takes three steps whether using Bracha's broadcast~\cite{bracha1984} or the AVID broadcast~\cite{avid}. The state-of-the-art ABA construction, the Cobalt ABA~\cite{cobalt2018}, has three or four steps in each round and the protocol may terminate in one or several rounds.} The situation is in sharp contrast to partially synchronous BFT protocols which usually have fewer steps (PBFT~\cite{castro2002practical}, for instance, terminates in three steps in the worst-case scenario).

This paper proposes \mib, faster asynchronous BFT protocols with \textit{suboptimal resilience}, including \miba (the case of $n \geq 5f+1$) and \mibb (the case of $n \geq 7f+1$). 
Our technique is \textit{generic} and can be applied to \textit{all} of the four state-of-the-art asynchronous BFT protocols (HoneyBadgerBFT, BEAT, EPIC, and Dumbo). To illustrate our approach, we use BEAT as the underlying protocol, as BEAT is simpler than EPIC and Dumbo, and has the most efficient open-source implementation available~\cite{beatimplementation}. (In fact, \mib can be based on any ACS instantiation or any asynchronous BFT using RBC and ABA.) 
For both \miba and \mibb, each epoch may terminate in as few as just \textit{three} steps in the best-case scenario. 

\heading{\mib techniques in a nutshell.} At the core of \mib are new RBC constructions and new ABA combinations with suboptimal resilience, enabling faster termination and higher throughput. 

For \miba, we devise \mbc, an erasure-coded version of Imbs and Raynal's RBC (IR RBC)~\cite{Tradingofft2016} which terminates in two steps by requiring $n \geq 5f+1$. \mbc is bandwidth-efficient and step-optimal. In contrast, previous asynchronous BFT protocols use Bracha's broadcast or the AVID broadcast~\cite{avid}, either of which completes in three steps. \mbc integrates the technique of AVID (using Merkle tree) and we formally prove the correctness of \mbc. 
For \miba, we instantiate the ABA construction in the ACS framework using a new combination of Bosco's weakly one-step ABA (W1S)~\cite{song2008bosco} that requires $n \geq 5f+1$ and the Cobalt ABA~\cite{cobalt2018}. In ideal situations, the new ABA construction terminates in just one step; otherwise, it falls back to the state-of-the-art Cobalt ABA. 

For \mibb, considering the step-optimal property of \mbc, one may intuitively use \mbc, even if there are more than $7f+1$ replicas. We show that we can achieve better performance by asking a fraction of replicas to be passive learners rather than active RBC participants. Our new RBC construction, \mbc with learners, or simply \mbcl, involves (much) fewer messages.  We formally prove the correctness of \mbcl. 
For ABA, we combine Bosco's strongly one-step ABA (S1S) that requires $n \geq 7f+1$ and the Cobalt ABA.  

\heading{A (powerful) programming and evaluation platform.} We build an asynchronous BFT programming and evaluation platform that fulfills two goals. First, the platform allows us to answer important research questions. For instance, how could one be certain that our RBC and ABA choices would perform as expected? More importantly, do we have other combinations that would lead to BFT with better performance? The platform allows mixing and matching different RBC and ABA primitives. Our framework has a flexible but unified API and is highly expressive: in total, we design and implement seven fully-fledged asynchronous BFT instances using different RBC and ABA combinations.      

Second, the platform allows us to perform experiments under failures and attacks. To the best of our knowledge, while state-of-the-art asynchronous BFT protocols, including HoneyBadgerBFT, BEAT, EPIC, and Dumbo, claim that they are more robust than partially synchronous BFT protocols in failure scenarios, no experiments are provided to validate the claims. The only framework we know does so is RITAS~\cite{ritas2008} that perform evaluation for one specific asynchronous BFT protocol using less than 10 replicas in LANs. Our platform can, however, perform experiments under various failure scenarios (crash, Byzantine, attacks) and allow us to compare different BFT protocols in a unified framework in a systematic manner.

\heading{Our contributions.} We summarize our contributions in the following: 

\begin{packeditemize}
	\item We design new distributed system primitives with suboptimal resilience, including new RBC constructions and ABA combinations. In particular, we provide an erasure-coded version of IR RBC using Merkle tree and provide a learner-version of RBC (where some replicas are passive learners). We formally prove the correctness of the new RBC constructions.
	\item We build a highly flexible \mib framework allowing  mixing and matching different RBC and ABA primitives. The framework consists of asynchronous BFT protocols described (\miba and \mibb) and five other variants. We provide meaningful tradeoffs among various situations, echoing the well-known claim that there is no ``one-size-fits-all" BFT.	
	\item We design experiments for asynchronous BFT protocols in failure and attack scenarios. This is the first systematic evaluation for these recent asynchronous BFT protocols using the ACS framework.   
	\item We have evaluated all seven \mib protocols and their competitors on Amazon EC2 with hundreds of experiments using up to 140 instances. We show that almost all \mib instances, in particular, \miba and \mibb, are much more efficient, in terms of both latency and throughout than their asynchronous BFT counterparts. Moreover, we show existing asynchronous BFT protocols, not just \mib protocols, are indeed robust against failures and attacks. We report many interesting results for different scenarios. 
		
\end{packeditemize}

\section{Related Work}\label{relatedwork}

\heading{BFT with suboptimal resilience.} A number of BFT protocols assume $n>3f+1$. For instance, Q/U requires $5f+1$ replicas to tolerate $f$ failures and achieves fault-scalability that tolerates increasing numbers of failures without largely decreasing performance~\cite{abd2005QU}.
BChain5 uses $5f+1$ replicas to simplify
the failure detection mechanism and remove the need for replica reconfiguration~\cite{bchain}. 
Zyzzyva5 uses $5f+1$ replicas and trades the number of replicas in the system against performance in the presence of faults~\cite{zyzzyva}. 
FaB Paxos \cite{martin2004fast} is efficient partially synchronous BFT protocol using $5f+1$ replicas and having 3 communication steps per request.

\heading{Efficient asynchronous atomic broadcast and BFT.} Most of the efficient asynchronous atomic broadcast (BFT) protocols follow the Ben-Or's ACS framework~\cite{ben1983another}, including SINTRA~\cite{sintra}, HoneyBadgerBFT~\cite{miller2016honey}, BEAT~\cite{duan2018beat}, EPIC \cite{EPIC2020}, and Dumbo~\cite{Dumbo2020}. They are different only in concrete instantiations. SINTRA, HoneyBadgerBFT, BEAT, and Dumbo achieve static security, where the adversary needs to choose the set of corrupted replicas before the execution of the protocol. In contrast, EPIC attains stronger adaptive security, 
where the adversary can choose to corrupt replicas at any moment during the execution of the protocol. 
Dumbo devises a new way of instantiating the ACS framework by using fewer ABA instances and achieves better performance. 
There are, however, efficient asynchronous BFT protocols that do not follow the ACS framework, including, for instance, RITAS~\cite{ritas2008}. DBFT~\cite{DBFT2018} relies on an asynchronous framework but works in partially synchronous environments and is very efficient.

\heading{Asynchronous binary agreement (ABA) with optimal resilience.}
Beginning with Ben-Or~\cite{ben1983another} and Rabin~\cite{rab83}, a significant number of ABA protocols have been proposed~\cite{bracha1987,cobalt2018,crabin1993,cachin2005random,tou84,rab83,bg1993,fmr2005,mhr14,sr2008,st1987,zielinski2006optimistically}. Cachin, Kursawe, and Shoup (CKS)~\cite{cachin2005random} proposed an ABA with optimal resilience and $O(n^2)$ message complexity. 
Mostefaoui, Moumen, and Raynal (MMR)~\cite{mhr14} proposed the first signature-free ABA with the same message complexity as the CKS ABA~\cite{cachin2005random}. 
The MMR ABA protocol was used by HoneyBadgerBFT, BEAT, and Dumbo. 
It was later reported that the MMR ABA has a liveness issue when being instantiated using
any coin-flipping protocols known~\cite{BCbug}. 
The Cobalt ABA protocol resolves the issue at the price of one more step for each round~\cite{cobalt2018}. The Cobalt ABA is used in EPIC and an open-source implementation of BEAT~\cite{beatimplementation}. Dumbo recently updated its ePrint version~\cite{dumboeprint20} by using the Cobalt ABA (code still unavailable). 

\heading{ABA with suboptimal resilience.}
Assuming $n\geq5f+1$, Berman and Garay~\cite{bg1993} presented a common coin based ABA protocol that has only two steps in each round. The ABA protocol by Friedman, Mostefaoui, and Raynal (FMR)~\cite{pcr2009} extended BG and reduced the number of steps within a round to one. Song and van Renesse~\cite{song2008bosco} proposed Bosco that terminates in one step in ideal situations, a protocol that we use in this paper.

\section{System and Threat Model}\label{system}

We consider a Byzantine fault-tolerant state machine replication (BFT) protocol with $n$ replicas, at most $f$ of which may exhibit arbitrary behavior (Byzantine failures). 
Most of BFT protocols assume optimal resilience with $n \geq 3f+1$. 
In this work, we consider suboptimal resilience with $n \geq 5f+1$ and $n \geq 7f+1$.

In BFT, replicas deliver transactions (requests) submitted by clients and send replies to clients.   
A BFT protocol should satisfy the following properties: 

\begin{packeditemize}

\item \textbf{Agreement}: If any correct replica delivers a transaction $tx$, then every correct replica delivers $tx$.

\item \textbf{Total order}: If a correct replica has delivered transactions $\langle tx_0,tx_1,\cdots,tx_j \rangle$ and another has delivered $\langle tx_0^{'},tx_1^{'},\cdots,tx_{j^{'}}^{'} \rangle$, then $tx_i=tx_i^{'}$ for $0 \leq i\leq min(j,j^{'})$.

\item \textbf{Liveness}: If a transaction $tx$ is submitted to $n-f$ replicas, then all correct replicas will eventually deliver~$tx$.

\end{packeditemize}

According to the timing assumptions, BFT protocols can be divided into three categories: \textit{asynchronous},  \textit{synchronous},
or \textit{partially synchronous}~\cite{dwork1988consensus}. Asynchronous BFT systems
make no timing assumptions on message processing or transmission
delays. Synchronous BFT systems have a known bound on message processing delays and
transmission delays.
Partially synchronous BFT systems lie in-between: messages will be delivered within a time bound, but the bound may be
unknown to anyone.
Asynchronous BFT protocols are inherently more robust than other BFT
protocols. This paper considers purely asynchronous systems making no timing assumptions on message processing or transmission delays.

\section{Building Blocks}
This section reviews the building blocks for \mib. 

\heading{Erasure coding.} An $(m,n)$ maximum distance separable (MDS) erasure coding scheme can encode $m$ data blocks (fragments) into $n$ ($n\geq m$) coded blocks, and all blocks can be recovered from any $m$-size subset of coded blocks via a decode algorithm. This paper by default uses MDS erasure coding.

\heading{Byzantine reliable broadcast (RBC).}
In RBC, a sender broadcasts a message to all other replicas in a group. 
An asynchronous RBC protocol satisfies the following properties: 

\begin{packeditemize}

	\item  \textbf{Validity}: If a correct replica $p$ broadcasts a message $m$, then $p$ eventually delivers $m$.  
	
	\item \textbf{Agreement}: If some correct replica delivers a message $m$, then every correct replica eventually delivers $m$.

	\item \textbf{Integirty}: For any message $m$, every correct replica delivers $m$ at most once. Moreover, if the sender is correct, then $m$ was previously broadcast by the sender.
	
\end{packeditemize}

\heading{Asynchronous binary agreement (ABA).}
In ABA, each replica has a binary value $v\in\{0,1\}$ (a vote) as the input, and correct replicas eventually deliver the same binary value as the output. ABA guarantees the following properties.

\begin{packeditemize}
	
	\item \textbf{Validity}: If a correct replica delivers a value $v$, the $v$ was proposed by at least one correct replica. 
	
	\item \textbf{Agreement}: If a correct replica delivers $v$ and another correct replica delivers $v'$, then $v=v'$. 
	
	\item \textbf{Termination}: All correct replicas eventually deliver a value with probability 1.
	
	\item \textbf{Unanimity}: If all correct replicas input the same initial value $v$, then a correct replica delivers $v$.
	
\end{packeditemize}

ABA protocols proceed in rounds, each of which includes several steps. We define one-step ABA as one ensuring one-step communication under the unanimity property, i.e., replicas terminate in one step if all correct replicas propose the same binary value~\cite{song2008bosco}.

\section{Technical Overview}\label{overview}

\begin{figure}[t!]
	\vspace{-6pt}
	\centering
	\includegraphics[width=0.8\linewidth]{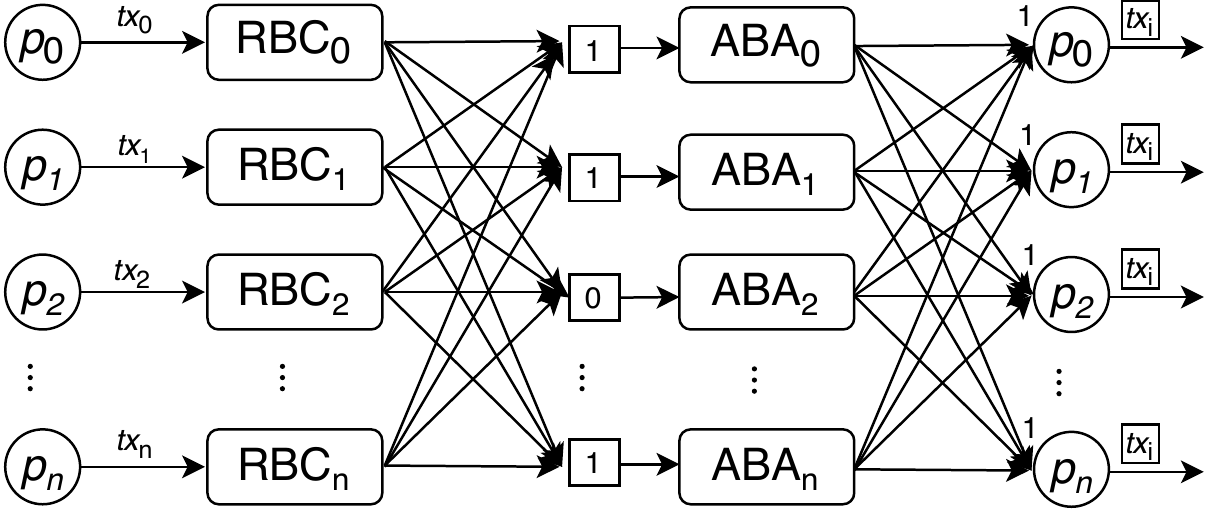}
	\caption{The ACS consensus workflow.}
	\label{fig:framework}
	\vspace{-12pt}
\end{figure}

State-of-the-art asynchronous BFT protocols, such
as HoneyBadgerBFT, BEAT, EPIC, and Dumbo, follow the asynchronous common subset (ACS) framework which includes a RBC phase and an ABA phase. 

We consider asynchronous BFT protocols (collectively called \mib) using the ACS framework with \textit{suboptimal resilience}.
At the core of the \mib protocols are new RBC constructions and new ABA combinations with suboptimal resilience. These new RBC constructions and ABA combinations are rather generic and can be applied to any ACS instantiations. 
For illustrative purposes, we follow the BEAT workflow (as depicted in~\figref{fig:framework}): BEAT is slightly simpler than EPIC and Dumbo and has the most efficient open-source implementation available~\cite{beatimplementation}. 

\heading{ABA with suboptimal resilience.}
HoneyBadgerBFT, BEAT, and Dumbo used the MMR ABA with optimal resilience~\cite{mhr14}. The MMR ABA 
includes two to three steps in each round.
The protocol, however, was found to have a liveness issue~\cite{BCbug}. 
The Cobalt ABA protocol resolves the problem at the price of one more step for each round~\cite{cobalt2018}. EPIC uses the Cobalt ABA, and some asynchronous BFT libraries (e.g., BEAT~\cite{beatimplementation}) have updated their implementation using the Cobalt ABA. 

For \mib, we use Bosco's one-step ABA protocols \cite{song2008bosco}: weakly one-step ABA (W1S) using $n \geq 5f+1$ replicas and strongly one-step ABA (S1S) using $n \geq 7f+1$ replicas. Both W1S and S1S terminate in as minimum as one single step (one round). W1S achieves this property when all replicas propose the same binary input (contention-free) and there are no faulty replicas (failure-free). S1S achieves this property under the contention-free condition but does not assume a failure-free condition. Bosco's ABA needs to run a backup ABA protocol when the conditions are not satisfied. We thus use a combination of W1S and the Cobalt ABA for \miba (the case of $n \geq 5f+1$) and a combination of S1S and the Cobalt ABA for \mibb (the case of $n \geq 7f+1$).   

\heading{RBC with suboptimal resilience.}
We devise \mbc, an erasure-coded version of IR RBC~\cite{Tradingofft2016} which terminates in two steps by requiring $n \geq 5f+1$. \mbc is bandwidth-efficient and step-optimal. 
The low bandwidth property has shown to be extremely useful for the performance of HoneyBadgerBFT and more so for BEAT; it turns out that the step-optimal property also improves the system performance.

\heading{RBC with learners.} 
When $n \geq 5f+1$, we have already obtained a step-optimal RBC (MBC) which terminates in two steps. There does not exist a one-step RBC, no matter how many replicas one uses. While one may consider simply using MBC directly for the case of $n \geq 7f+1$, we can actually do better. 

We use the concept of \textit{learners} (see, e.g., Paxos~\cite{Paxos2001}) and propose \textit{RBC with learners}. 
If there is a RBC that requires $n_1$ replicas and an ABA that requires $n_2$ replicas, where $n_2>n_1$, then $n_1$ replicas are active replicas and $n_2-n_1$ replicas are learners. The learners do not actively participate in RBC but only learn the results. When an active replica delivers a message, it forwards the message to all learners. A learner delivers a message when it receives $n_1-f$ matching messages and then enters the ABA phase. Compared to regular RBC, RBC with learners reduces the number of messages transmitted. 
RBC with learners is a general primitive. 
The specific MBC extension is called \mbcl, and the AVID extension is called \avidl. 

\heading{The \mib framework.} 
To examine if the above new primitives or combinations could improve performance as expected, we build a highly modular and expressive framework. In such a framework, we mix and match various RBC and ABA primitives to design and implement seven asynchronous BFT protocols with suboptimal resilience. 
The framework is modular: various components are programmed to fit in a unified but flexible standard API (for varying $f$ and $n$'s). It is the framework that allows us to have a clear picture on the performance bottlenecks for all \mib instances, validate our theoretical design, and help find meaningful trade-offs.

\section{\mib}\label{protocol}


\subsection{\mib framework}

This section describes the \mib protocols. 
In \mib, we propose new RBC primitives and new ABA combinations as building blocks. The building blocks can be applied to all asynchronous BFT systems using the ACS framework or any asynchronous BFT using RBC and ABA. To illustrate our approach, \mib is built on top of BEAT which has the most efficient open-source implementation available~\cite{beatimplementation}.

\mib protocols are different only in concrete RBC and ABA instantiations. So when describing \mib in general, we use RBC and ABA in a black-box manner. 
\figref{fig:mainalg} depicts the \mib framework which is the same as BEAT. 
\mib proceeds in epochs numbered by $r$ (initially, $0$).
In an epoch, replicas choose a subset of transactions as a \textit{proposal} from
their transaction pool and agree on a set containing the union of the proposals of at least $n-f$ replicas.
We define $B$ as the batch size of the transactions for an epoch; the batch size for a replica $b =  \lceil B/n \rceil$. 
Replicas first run a RBC phase
to broadcast their proposals. 
Then they run an ABA phase, where $n$ parallel ABA instances are invoked. The $i$-th ABA instance agrees on whether the proposal of replica $p_i$ has been delivered in the RBC phase. If a correct replica $p_j$ terminates the $i$-th ABA instance with $1$, the proposal from $p_i$ is delivered. Otherwise, the proposal is not included. 
We ensure that at least $n-f$ ABA instances terminate with $1$ and the union of the transactions from at least $n-f$ replicas are delivered. 
To do this, each replica abstains from proposing $0$
until $n-f$ ABA instances have been delivered by the replica. 
As in BEAT, \mib uses threshold encryption to avoid transaction censorship and achieve liveness (the pseudocode of threshold encryption is not shown in~\figref{fig:mainalg}).

\begin{figure}[t!] \small
	\fbox{\parbox{3.3in}{\vspace{0ex}
			\parbox{3.3in}{
    			\phantom{00000000000}\textbf{Algorithm: \mib Protocol (Replica $p_i$)}  \\
				\textbf{Initialization} \\
				\phantom{0} let $B$ be the batch size parameter \\
				\phantom{0} let $\mathsf{buf}$ $\leftarrow \emptyset$ be a transaction buffer \\
				\phantom{0} let \{RBC$_j$\}$_{j\in n}$ and \{ABA$_j$\}$_{j\in n}$ be the $j$-th instance for \\
				\phantom{0} RBC$_j$ and ABA$_j$\\
				\phantom{0} let $output \leftarrow \emptyset$ be the output buffer\\
				\phantom{0} let $r \leftarrow 0$ be the epoch number\\
				\textbf{epoch $\emph{r}$}\\
				\phantom{0} select $b=\lceil B/n \rceil$ random transactions from the first $B$  \\
				\phantom{0} elements in $\mathsf{buf}$ as a proposal $value$\\
				\phantom{0} \textbf{upon} input value $value$ \\
				\phantom{000} input $value$ to RBC$_i$ \\
				\phantom{0} \textbf{upon delivery} of $value_j$ from RBC$_j$\\
				\phantom{000} \IF ABA$_j$ has not yet been provided input, input $1$ to ABA$_j$\\
				\phantom{0} \textbf{upon delivery} of $1$ from ABA$_j$ and $value_j$ from RBC$_j$\\
				\phantom{000} $output \leftarrow output \cup value_j$\\
				\phantom{0} \textbf{upon delivery} of $1$ from at least $n-f$ ABA instances\\
				\phantom{000} \textbf{for} each ABA$_j$ instance that has not been provided input\\
				\phantom{00000} input $0$ to ABA$_j$ \\
				\phantom{0} \textbf{upon termination} of all the $n$ ABA instances\\
				\phantom{000} \textbf{deliver} $output$ \\
				\phantom{0} $r \leftarrow r+1$   
			}
	}}
	\caption{The \mib algorithm for $p_i$. }
	\label{fig:mainalg}
\end{figure}

\subsection{\miba}

For \miba, we instantiate the ABA component in the ACS framework using a new combination of Bosco's weakly one-step ABA (W1S)~\cite{song2008bosco} that requires $n \geq 5f+1$ and the Cobalt ABA~\cite{cobalt2018}.
We also devise \mbc, an erasure-coded version of IR RBC~\cite{Tradingofft2016} which completes in two steps (optimal) and requires $n \geq 5f+1$.

\begin{figure}[h!] \small
	\centering
	\fbox{\parbox{3.3in}{\vspace{0ex}
			\parbox[b]{3.3in}{
    			\phantom{000000000000000}\textbf{Algorithm: W1S/S1S}  \\
			    \textbf{Initialization}
			    \\
				\phantom{0} $r \leftarrow 0$ \hfill \{round\}\\
				\phantom{0} $v_{p}$ \hfill \{input value\}\\
				\textbf{round $\emph{r}$}  \\
				\phantom{0} broadcast \bval{v_p}\hfill \{broadcast input\}\\
				\phantom{0} \upon \bval{v} from $n-f$ replicas \\
				\phantom{00}\textbf{if} more than $\lceil(n+3f)/2\rceil$~\bval{v} messages contain the same\\
				\phantom{0000}value $v$ \\
				\phantom{0000}\textbf{deliver} $v$ \hfill \{terminate the protocol\}\\
				\phantom{00}\textbf{if} more than $\lceil(n-f)/2\rceil$~\bval{v} messages contain the same \\
				\phantom{0000}value $v$, and there is only one such value $v$\\
				\phantom{0000}$v_{p} \leftarrow v$\\
				\phantom{0} \textbf{backup-ABA($v_p$)}
	}}}
	\caption{The algorithm for W1S and S1S.}
\label{fig:Bosco}

\end{figure}

\heading{Weakly one-step (W1S).} The state-of-the-art ABA for the $n \geq 3f+1$ case, the Cobalt ABA, requires at least three steps in each round. For the case of $n \geq 5f+1$, we use the Bosco's weakly one-step ABA protocol (W1S)~\cite{song2008bosco}.

W1S guarantees that if there are \textit{no faulty} replicas and all replicas propose the \textit{same} initial value $v$, then all the correct replicas deliver $v$ and terminate the protocol in one step (one round). \figref{fig:Bosco} describes the pseudocode of W1S protocol. Each replica has a binary input $v_p$ to the ABA. In the first step, each replica $p_i$ broadcasts a message \bval{v_p}. A replica $p_i$ waits for \bval{v} from $n-f$ replicas (including itself). If more than $\lceil \frac{n+3f}{2} \rceil$ \bval{v} messages include the same value $v$, a replica delivers $v$ and terminates the protocol. 
If more than $\lceil \frac{n-f}{2}\rceil$ \bval{v} messages include the same $v$, a replica sets its local value to $v_p=v$. If the replica does not deliver any value in one step, it invokes a backup ABA protocol. In \miba, we use the Cobalt ABA protocol~\cite{cobalt2018} as the backup ABA protocol.

\begin{figure}[t]
	\vspace{-6pt}
	\centering
	\includegraphics[width=0.72\linewidth]{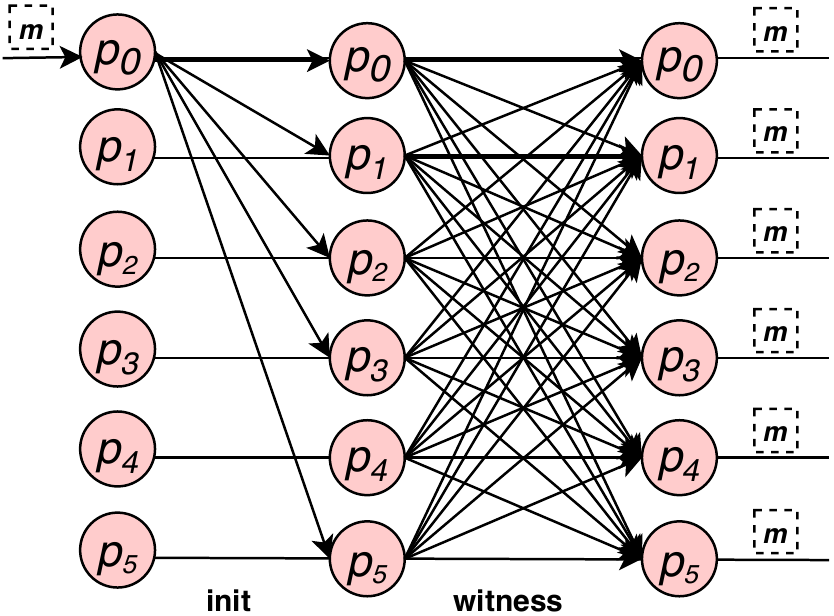}
	\caption{The \mbc workflow, where $p_0$ broadcasts a message $m$ and there are no faulty replicas.}
	\label{fig:mbc}
\end{figure}

\heading{MBC}.
\mbc is an erasure-coded version of IR RBC~\cite{Tradingofft2016} which completes in two steps. \mbc is bandwidth-efficient and step-optimal. We show the \mbc workflow in \figref{fig:mbc} and the \mbc pseudocode in \figref{fig:Damien}.

\begin{table}[t]
\centering
\begin{tabular}{|c||c|c|c|}
\hline
~ & resilience level &  steps & number of messages \\
\hline\hline
\avid        &   $n \geq 3f+1$              &   3 & $2n^2+n$
\\ 
\hdashline[1pt/1pt]                 
\mbc         &   $n \geq 5f+1$              &   2  & $n^2+n$\\ 
\hline
                       
\end{tabular}
\vspace{2pt}
\caption{Comparison of the RBC algorithms.}
\label{table:twoalgorithms}
\vspace{-1em}
\end{table}

\begin{figure}[!ht] \small
	\centering
	\fbox{\parbox{3.3in}{\vspace{0ex}
			\parbox[b]{3.3in}{
			\phantom{00000000000000}\textbf{Algorithm: \mbc (Replica $p_i$)} \\
			\textbf{upon}~input($m$)\hfill \{if $p_{sender}=p_i$\}\\
			\phantom{0000} let $l_{j}$ be the $j$-th block of $(n-2f,n)$ erasure coding \\
			\phantom{0000} scheme applied to $m$\\
			\phantom{0000} $h$ is the root of the Merkle tree for the $\{l_{j}\}_{j\in[0..n-1]}$ blocks\\
			\phantom{0000} send $\mathsf{init}(h,b_{j},l_{j})$ to each $p_j$, where $b_j$ is the $j$-th Merkle\\
			\phantom{0000}  tree branch\\
			
			\textbf{upon}~\textbf{receiving}~$\mathsf{init}(h,b_{i},l_{i})$ from $p_{sender}$\\
			\phantom{0000} broadcast $\mathsf{witness}(h,b_{i},l_{i})$\\
			
			\textbf{upon}~\textbf{receiving} $n-2f$ valid~$\mathsf{witness}(h,b_{j},l_{j})$ \\
		    \phantom{00000}interpolate $\{l_j'\}$ from the $n-2f$ blocks\\
		    \phantom{00000}recompute Merkle root $h'$ and if $h'\neq h$ then abort\\
			\phantom{00000}if $\mathsf{witness}(h,b_{i},l_{i})$ is not sent \\
			\phantom{0000000}broadcast $\mathsf{witness}(h,b_{i},l_{i})$\\

	 \textbf{upon} \textbf{receiving} $n-f$ valid $\mathsf{witness}(h,b_{j},l_{j})$\\
			\phantom{00000}$m$ $\leftarrow$ decode($\{l_j\}$) \hfill \{from any $n-2f$ blocks\}\\
			\phantom{00000}\textbf{deliver}($m$)
	}}}
	\caption{The \mbc protocol. A message $(h, b_j, l_j)$ is valid, if $b_j$ is a valid Merkle tree branch for the Merkle tree root $h$ and the data block $l_j$.}
\label{fig:Damien}
\end{figure}

As depicted in~\figref{fig:Damien},
to broadcast a message $m$, a replica $p_i$ applies the $(n-2f, n)$ erasure coding scheme to generate $n$ blocks, where the $j$-th block is denoted as $l_j$. The replica $p_i$ then generates a Merkle tree for the $n$ blocks. Finally, for $j$ between $0$ to $n-1$, replica $p_i$ sends an $\mathsf{init}(h, l_j, b_j)$ message to the $j$-th replica $p_j$, where $h$ is the root of the Merkle tree and $b_j$ is the $j$-th Merkle tree branch. We say a message $(h, b_j, l_j)$ is valid, if $b_j$ is a valid Merkle tree branch for the Merkle tree root $h$ and the data block $l_j$.

If a replica $p_i$ receives an $\mathsf{init}(h,l_i, b_i)$ message, $p_i$ broadcasts $\mathsf{witness}(h,l_i, b_i)$.  

If a replica $p_i$ receives $n-2f$ valid $\mathsf{witness}(h,l_j, b_j)$ messages,
$p_i$ interpolates all $n$ blocks from $n-2f$ blocks, recomputes the Merkle tree root $h'$. If $h'\neq h$ and it has not broadcast any $\mathsf{witness()}$ message, it broadcasts $\mathsf{witness}(h,l_i, b_i)$. Otherwise, it simply aborts.  

If $p_i$ receives $n-f$ valid $\mathsf{witness()}$ messages, $p_i$ recovers the original input $m$ and delivers $m$.

As shown in Table~\ref{table:twoalgorithms}, assuming the same $n$, \mbc has fewer steps and fewer messages than \avid (used in HoneyBadgerBFT, BEAT, and Dumbo).

\begin{theorem}
	The MBC protocol in~\figref{fig:Damien} is a reliable broadcast protocol.
\end{theorem}

\heading{Proof:}
	We prove the theorem from scratch (instead of using the proof of IR MBC in a black-box manner).

We first prove validity. If a correct sender broadcasts a message $m$, it will erasure codes the message $m$ into $n$ blocks $\{l_{j}\}_{j\in[0..n-1]}$ (where $n-2f$ blocks are sufficient to recover $m$), and generates a Merkle tree proof $(h, b_j)$ for each block $l_j$. Then the sender sends $j$-th block $l_j$ and the corresponding proof $(h, b_j)$ to the corresponding replica $p_j$. Upon receiving an $\mathsf{init}$ message, each replica will verify whether the message is valid and then broadcasts the $\mathsf{witness}$ messages to all replicas. Eventually, all correct replicas will receive $n-f$ $\mathsf{witness}$ valid messages. Since the sender is correct, the recomputed Merkle root must equal the agreed one. Any correct replica $p_j$ can recover $m$ using $n-2f$ erasure coding scheme with matching root and then deliver $m$. 

We now prove agreement. If some correct replica $p_i$ delivers a message $m$ with some $h$, then the replica must have received $n-f$ valid $\mathsf{witness}$ messages with the matching root $h$. Among these $n-f$ replicas, at least $n-2f$ replicas are correct. These correct replicas must have received the $\mathsf{init}(h, \cdot, \cdot)$ messages and must have sent $\mathsf{witness}(h, \cdot, \cdot)$ messages to all replicas. Therefore, all correct replicas will receive valid $n-2f$ $\mathsf{witness}(h, \cdot, \cdot)$ messages. We claim that if $p_i$ delivers a message with $h$, then except with negligible probability, any other correct replica will not abort (as the recomputed Merkle root $h' = h$). Otherwise, one can find an adversary attacking the Merkle tree (more concretely, attacking the collision resistance property of the underlying hash function of the Merkle tree). Therefore, all correct replicas will broadcast $\mathsf{witness}(h, \cdot, \cdot)$ and eventually all correct replicas will receive $n-2f$ valid $\mathsf{witness}(h, \cdot, \cdot)$ messages. Again, according to the property of the Merkle tree, all correct replicas can recover and deliver the same $m$.  

Finally, integrity holds by inspection of the protocol.
This completes the proof of the theorem. \hfill $\Box$

\subsection{\mibb}
In \mibb, we combine Bosco's strongly one-step ABA (S1S) for $n \geq 7f+1$ and the Cobalt ABA. We design a new RBC construction, \mbc with learners, or simply \mbcl, to further reduce the number of messages transmitted.

\heading{Strongly one-step (S1S)}. The strongly one-step ABA (S1S) is another one-step algorithm for $n \geq 7f + 1$ in Bosco~\cite{sr2008}. S1S runs the same algorithm as W1S but achieves different properties. If all correct replicas propose the same initial value $v$, all correct replicas deliver $v$ in one communication step. 
Namely, S1S guarantees one-step termination under contention-free situations; it does not require the failure-free conditions needed for the one-step termination in W1S. As in \miba, \mibb uses the Cobalt ABA as the backup ABA protocol.

\heading{\mbcl}. 
When $n \geq 5f+1$, \mbc is already a step-optimal RBC. While intuitively for the case of $n \geq 7f+1$, one could use \mbc directly, we actually use \textit{MBC with learners} (\mbcl). In such a primitive, some replicas are just learners instead of active replicas participating in the main broadcast process. But of course, \mbcl remains a standard RBC satisfying all RBC properties. 

\begin{figure}[h]
	\vspace{-6pt}
	\centering
	\includegraphics[width=0.7\linewidth]{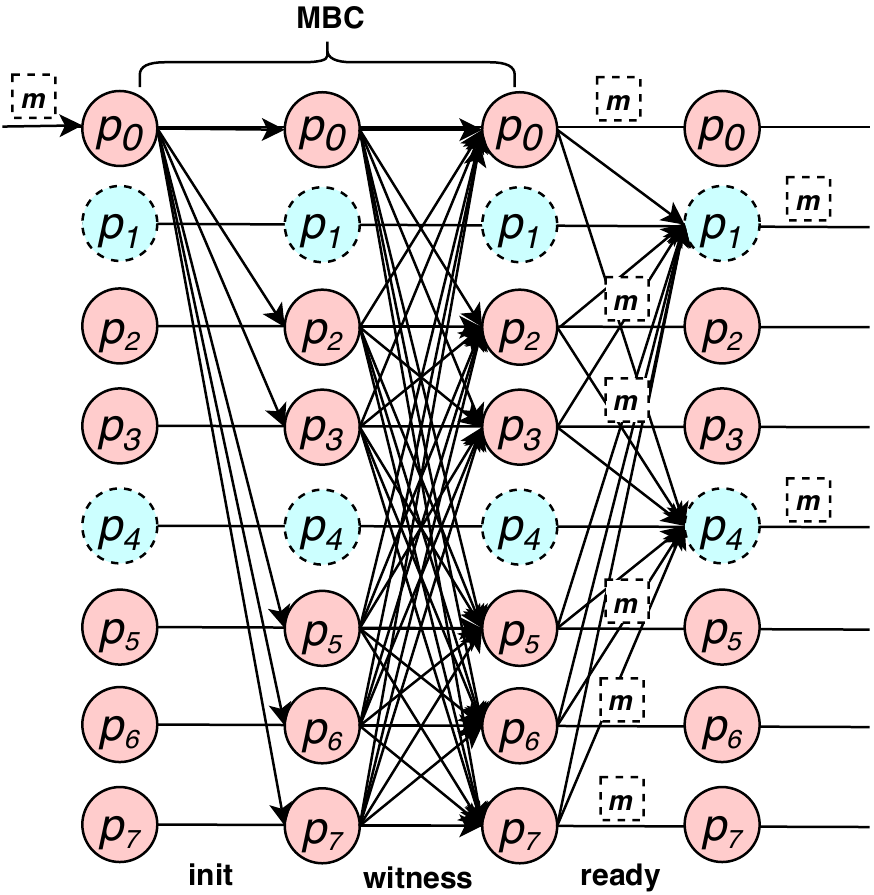}
	\caption{The \mbcl workflow. Replica $p_0$ is the broadcaster, replicas in solid circles are active replicas, and replicas in dashed circles are learners. Active replicas run the \mbc protocol to deliver the message $m$ and forward it to the learners once the message is delivered.}
	\label{fig:mbcl}
\end{figure}

\begin{figure}[h!] \small
	\centering
	\fbox{\parbox{3.3in}{\vspace{0ex}
			\parbox[b]{3.3in}{
			\phantom{0000000000}\textbf{Algorithm: \mbcl in \mibb (Replica $p_i$)} \\
			\textbf{upon}~input($m$)\hfill \{if $p_{sender}=p_i$\}\\
			\phantom{0000} $active\_rep$ $\leftarrow$ $n_1=5f+1$ active replicas \\
			\phantom{0000} let $l_{j}$ be the $j$-th block of $(n_1-2f,n_1)$ erasure coding \\
			\phantom{0000} scheme applied to $m$\\
			\phantom{0000} $h$ is the root of the Merkle tree for the $\{l_j\}_{j\in[0..n_1]}$ blocks\\
			\phantom{0000} send $\mathsf{init}(h,b_{j},l_{j},active\_rep)$ to each $p_j$, where $b_j$ is \\
			\phantom{0000} the $j$-th Merkle tree branch\\
			
			\textbf{upon}~\textbf{receiving}~$\mathsf{init}(h,b_{i},l_{i},active\_rep)$ from $p_{sender}$\\
			\phantom{0000} broadcast $\mathsf{witness}(h,b_{i},l_{i},active\_rep)$\\
			
		    \textbf{upon} \textbf{receiving} $n_1-2f$ valid $\mathsf{witness}(h,b_{j},l_{j},active\_rep)$ \\
		    \phantom{00000}interpolate $\{l_j'\}$ from the $n_1-2f$ blocks\\
		    \phantom{00000}recompute Merkle root $h'$ and if $h'\neq h$ then abort\\
		    \phantom{00000}if $\mathsf{witness}(h,b_{i},l_{i},active\_rep)$ has not been sent \\
		    \phantom{0000000}broadcast $\mathsf{witness}(h,b_{i},l_{i},active\_rep)$\\
		    
		    \textbf{upon}~\textbf{receiving} $n_1-f$ valid $\mathsf{witness}(h,b_{j},l_{j},active\_rep)$ \\
		    \phantom{00000}send $\mathsf{ready}(h,b_{i},l_{i})$ to $2f$ learners\\
			\phantom{00000}$m$ $\leftarrow$ decode($\{l_j\}$) \hfill \{from any $n_1-2f$ blocks\}\\
			\phantom{00000}\textbf{deliver}($m$)\\
		    
			\textbf{upon}~\textbf{receiving} $n_1-f$ valid $\mathsf{ready}(h,\cdot,\cdot)$ \hfill \{learners\}\\
			\phantom{00000}$m$ $\leftarrow$ decode($\{l_j\}$)\\
			\phantom{00000}\textbf{deliver}($m$) 
	}}}
	\caption{\mbcl in \mibb. A message $(h,b_{i},l_{i})$ is valid, if $b_j$ is a valid Merckle tree branch for the Merkle tree root $h$ and the data block $l_j$.}
\label{fig:MiBwithlearner}
\end{figure}

We show in \figref{fig:mbcl} and \figref{fig:MiBwithlearner} the workflow and pseudocode of \mbcl, respectively. Assuming $n = 7f+1$ replicas, we have $5f+1$ replicas are active replicas and the other $2f$ replicas are learners. (We describe the selection principle shortly.)
To broadcast a message, a broadcaster $p_i$ runs \mbc to broadcast its input $m$ among active replicas. When an active replica $p_j$ delivers a message, it broadcasts a $\mathsf{ready}$ message to all learners. When a learner receives at least $4f+1$ valid $\mathsf{ready}$ messages from active replicas, it recovers the value $m$ and delivers it. 

In the \mibb environment, all replicas are active in the ABA phase. Namely, each replica votes for $1$ for an ABA instance after it delivers a value in the RBC phase regardless of whether it is an active replica or a learner. Each replica waits until at least $n-f$ ABA instances terminate with 1 before invoking other ABA instances with 0 as input. 

Our approach is generic. 
If there is a RBC that requires $n_1$ replicas (e.g., tolerating $f_1 = \lfloor \frac{n_1-1}{3} \rfloor$ failures) and an ABA that requires $n_2$ replicas (e.g., tolerating $f_2 = \lfloor \frac{n_2-1}{5} \rfloor$ failures) where $n_2 > n_1$, then some $n_2-n_1$ replicas are designated as learners who do not actively participate in RBC but later participate in ABA. The system tolerates $f_2$ failures. The $n_1$ replicas need to run RBC to deliver messages. When one of the $n_1$ replicas delivers a message, it forwards the message to all learners. A learner delivers a message when it receives $n_1-f_2$ matching messages and then enters the ABA phase.

In the ACS framework, $n$ parallel RBC instances are run concurrently. 
The principle of selecting active replicas is pretty arbitrary, as long as the system designer takes into account load balancing. For instance, in a system with replicas $\{p_0, \cdots, p_{n-1}\}$, the system designer can ask $p_i$ $(i \in [0,\cdots,n-1])$ to deterministically select replicas $\{p_i,\cdots, p_{(i+5f)~\text{mod}~7f}\}$. One could ask replicas to randomly select $5f+1$ replicas among all replicas, but the strategy is no better than the above-mentioned deterministic strategy that enables strictly even load balancing when $n$ concurrent RBC instances are run.  

While \mbcl has one more step than \mbc, \mbcl in fact greatly reduces the number of messages in \mibb. As mentioned earlier, the total number of messages in \mbc is $n^2+n$. Assuming $n=7f+1$, the total number of messages for \mbc is $49f^2+21f+2$. In \mbcl we use for \mibb, there are $2f$ learners. Each \mbcl instance involves $n-2f$, $(n-2f)^2$, and $2f(n-2f)$ messages in the $\mathsf{init}$, $\mathsf{witness}$, and $\mathsf{ready}$ steps, respectively. The total number of messages for \mbcl is $35f^2+17f+2$. 

\begin{theorem}
MBC-L in~\figref{fig:MiBwithlearner} is a reliable broadcast protocol.
\end{theorem}

\heading{Proof:} Validity and integrity hold by inspection of the protocol. We focus on agreement, showing that if a correct replica $p_i$ delivers $m$ then a correct replica $p_j$ eventually delivers $m$. We distinguish several cases:

\headit{Case 1: both $p_i$ and $p_j$ are active replicas.} In this case, agreement follows trivially from that of MBC. 

\headit{Case 2: $p_i$ is an active replica and $p_j$ is a learner.} If $p_i$ delivers a message $m$ with some $h$, then according to the agreement property of the underlying MBC, all correct replicas will deliver $m$ with the same $h$. These replicas will broadcast $\mathsf{ready}$ messages and eventually all learners will receive $n_1-f$ valid $\mathsf{ready}$ messages. According to the property of the Merkle tree, all learners can recover and deliver the same $m$.

\headit{Case 3: $p_i$ is an learner and $p_j$ is an active replica.} As $p_i$ is an learner, it must have received $n_1-f$ valid $\mathsf{ready}$ messages with matching $h$. This means that at least $n_1-2f$ active replicas have delivered the corresponding $m$. Due to the agreement property of underlying MBC protocol, all active replicas will deliver $m$.  

\headit{Case 4:  both $p_i$ and $p_j$ are learners.} 
This is similar to Case~3. 
Since $p_i$ is an learner, it must have received $n_1-f$ valid $\mathsf{ready}$ messages with the matching $h$. Hence, at least $n_1-2f$ active replicas have delivered the corresponding message $m$. Due to the agreement property of underlying MBC protocol, all active replicas will deliver $m$. These replicas will send valid $\mathsf{ready}$ messages so that all learners will obtain $n_1-f$ $\mathsf{ready}$ messages. The learners and active replicas agree on the same $h$ and except with negligible probability, they will obtain the same $m$.

This completes the proof of the theorem. \hfill $\Box$

\subsection{Other \mib Variants}

We build a modular and expressive \mib framework, where we mix and match various RBC and ABA primitives to construct five additional asynchronous BFT protocols with suboptimal resilience.
We summarize all \mib protocols in Table~\ref{table:MIBsevenprotocols}, including two \miba variants (\mibaa and \mibab) and three \mibb variants (\mibba, \mibbb, and \mibbc). The framework allows us to validate our theoretical design and help
identify meaningful protocol trade-offs among various combinations. 

\begin{figure}[h]
	\vspace{-6pt}
	\centering
	\includegraphics[width=0.8\linewidth]{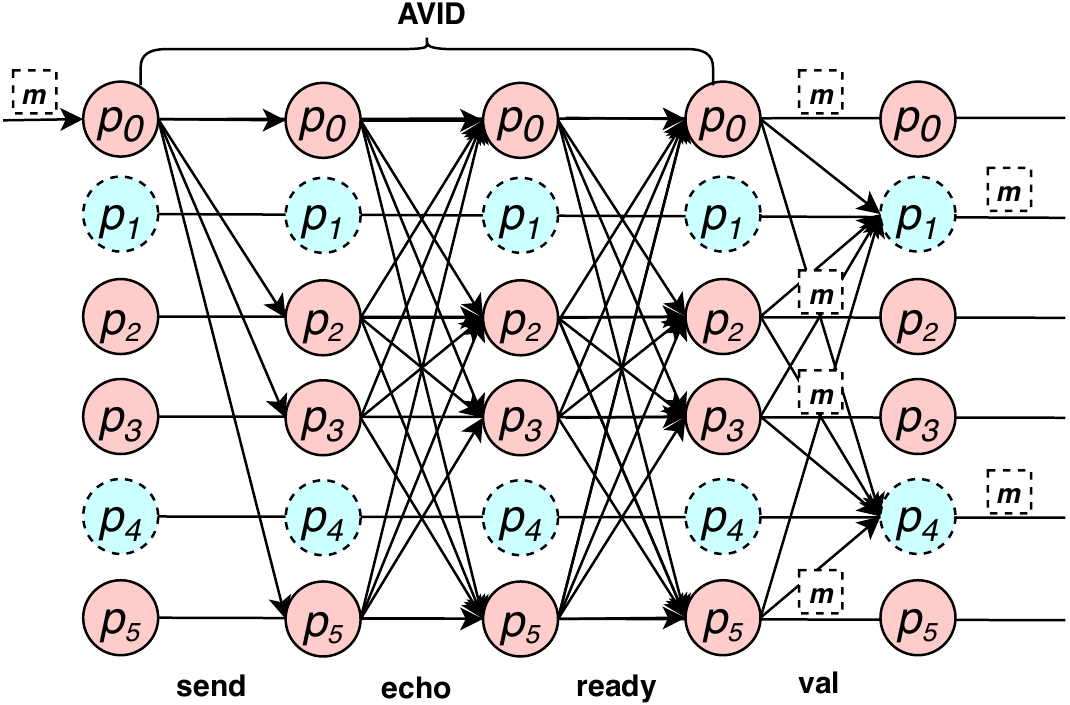}
	\caption{The \avidl workflow. $p_0$ is the broadcaster. Active replicas are denoted with solid circles and learners are represented in dashed circles.}
	\label{fig:avidl}
\end{figure}

Note the idea of RBC learners in \mbcl applies to the \avid broadcast for both the case of $5f+1$ and $7f+1$. We show in \figref{fig:avidl} and \figref{fig:Brachawithlearner} the workflow and pseudocode for the \avidl protocol that we use for \mibab. In \avidl, some $3f+1$ replicas are active replicas and the rest of replicas are learners. A broadcaster $p_i$ applies erasure coding to generate blocks for input $m$ and broadcasts each block to the corresponding replica. Active replicas run \mavid to deliver the input $m$. After the message is delivered, each active replica forwards learners the delivered message. A learner recovers the value $m$ using blocks received and delivers $m$. 
Compared to \mavid, \avidl has one additional step but less number of messages.

\begin{figure}[h!] \small
	\centering
	\fbox{\parbox{3.3in}{\vspace{0ex}
			\parbox[b]{3.3in}{
			\phantom{000000000}\textbf{Algorithm: \avidl in \mibab (Replica $p_i$)} \\
			\textbf{upon}~input($m$)\hfill \{if $p_{sender}=p_i$\}\\
			\phantom{0000} $active\_rep$ $\leftarrow$ $n_1=3f+1$ active replicas \\
			\phantom{0000} let $l_{j}$ be the $j$-th block of $(n_1-2f,n_1)$ erasure coding \\
			\phantom{0000} scheme applied to $m$\\
			\phantom{0000} $h$ is the Merkle tree root of $\{l_{j}\}_{j\in[0..n_1]}$ blocks\\
			\phantom{0000} send $\mathsf{send}(h,b_{j},l_{j},active\_rep)$ to each $p_j$, where\\
			\phantom{0000} $b_j$ is the $j$-th Merkle tree branch\\

			\textbf{upon}~\textbf{receiving}~$\mathsf{send}(h,b_{j},l_{j},active\_rep)$ from $p_{sender}$\\
			\phantom{0000} broadcast $\mathsf{echo}(h,b_{j},l_{j},active\_rep)$\\
			
			\textbf{upon}~\textbf{receiving}~$\mathsf{echo}(h,b_{j},l_{j},active\_rep)$ from $p_j$\\
			\phantom{0000} check if $b_j$ is a valid Merkle tree branch for $h$ and $l_j$\\
			\phantom{0000} check whether $i$ is in $active\_rep$ \\
			
			\textbf{upon}~\textbf{receiving} valid $\mathsf{echo}(h,\cdot,\cdot)$ from $n_1-f$ distinct parties\\
		    \phantom{00000}interpolate $\{l_j'\}$ from any $n_1-2f$ leaves received\\
		    \phantom{00000}recompute Merkle root $h'$ and if $h'\neq h$ then abort\\
		    \phantom{00000}if $\mathsf{read}(h)$ has not been sent\\
			\phantom{0000000}broadcast $\mathsf{ready}(h,active\_rep)$\\

			\textbf{upon}~\textbf{receiving} $f+1$~$\mathsf{ready}(h,active\_rep)$ from $p_{j}$\\
			\phantom{00000}if $\mathsf{ready(h)}$~has not yet been sent\\
			\phantom{0000000}broadcast~$\mathsf{ready}(h,active\_rep)$\\
			
			\textbf{upon}~\textbf{receiving} $2f+1$~$\mathsf{ready}(h,active\_rep)$ from $p_{j}$\\
			\phantom{00000}send $\mathsf{val}(h,b_{i},l_{i})$ to $2f$ learners\\
			\phantom{00000}$m$ $\leftarrow$ decode($\{l_j\}$)\\
			\phantom{00000}\textbf{deliver}($m$)\\
			
			\textbf{upon}~\textbf{receiving} $n_1-f$ valid $\mathsf{val}(h,\cdot,\cdot)$  \hfill \{learners\}\\
			\phantom{00000}$m$ $\leftarrow$ decode($\{l_j\}$)\\
			\phantom{00000}\textbf{deliver}($m$) 
	}}}
	\caption{\avidl in \mibab.}
\label{fig:Brachawithlearner}
\end{figure}

\begin{table}[h!]
\small
    \centering
	\begin{tabular}{|c||c|c|c|c|c|}
		\hline
		 & resilience level & ABA & RBC \\
		\hline\hline
		\miba &  $n \geq 5f+1$& W1S & \mbc \\
		\hdashline[1pt/1pt]
		\mibaa & $n\geq 5f+1$& W1S   & \avid \\
		\hdashline[1pt/1pt]
		\mibab  &$n\geq 5f+1$& W1S   & \avidl \\
		\hdashline[1pt/1pt]
		\mibb  &$n\geq 7f+1$  & S1S  & \mbcl  \\
		\hdashline[1pt/1pt]
		\mibba & $n\geq 7f+1$& S1S  & \avid \\
		\hdashline[1pt/1pt]
		\mibbb  &$n\geq 7f+1$ & S1S   & \mbc \\
		\hdashline[1pt/1pt]
		\mibbc  &$n\geq 7f+1$ & S1S   & \avidl \\
		
		\hline
	\end{tabular}
	 
	\vspace{3pt}
	\caption{MiB protocols. RBC with -L labels are protocols with learners. }
    \label{table:MIBsevenprotocols}
   
\end{table}

\section{Implementation and Evaluation}\label{sec:evaluation}

\heading{Implementation.} 
We build \mib from the open-source prototype of BEAT library \cite{beatimplementation} written in Python. We use the BEAT0 protocol (hereinafter BEAT for simplicity) as our baseline protocol.
The \mib programming framework is modular, with a unified API encompassing eight protocols---BEAT and all seven \mib protocols summarized in Table~\ref{table:MIBsevenprotocols}. 
We use the zfec library \cite{zfec} for erasure
coding. 
Following BEAT, for all \mib instances,
we use Shoup and Gennaro threshold encryption scheme~\cite{SecuringThreshold98}  and the CKS threshold PRF \cite{Randomoracles2005} as the threshold encryption and the coin-flipping protocol, respectively. We use the prime256v1 curve with 128-bit security for the above two threshold cryptographic primitives. All the crypto schemes are implemented using the Charm Python crypto library\cite{charm}.

\begin{figure*}[ht!]
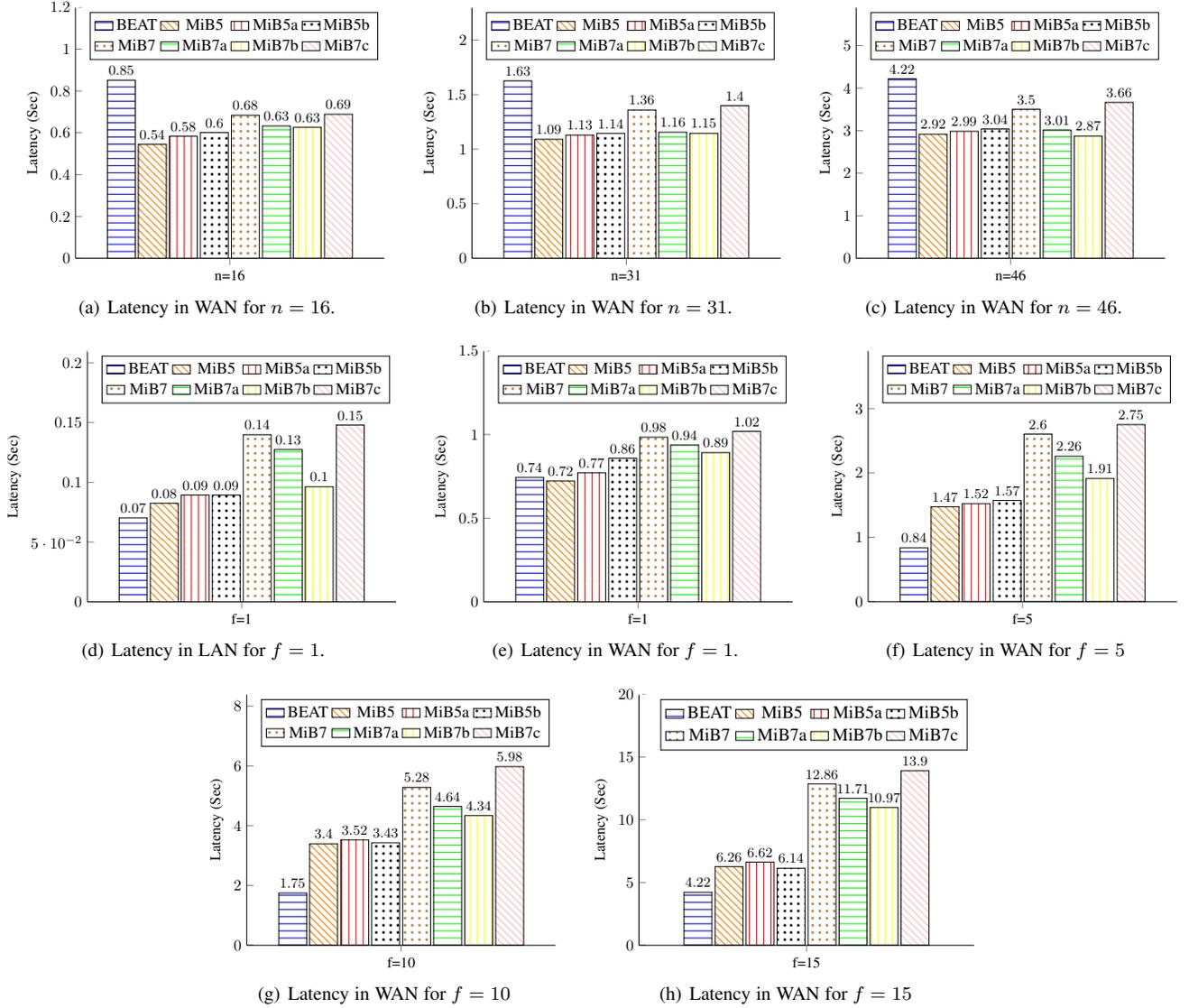

	\centering
	\hspace{0.5pt}
	\subfigure[Latency in WAN for $n=16$.]{ 
		\input{figs/MIB-latency_n=16}
		\label{fig:MIB-latency(n=16)}
	}
	\hspace{0.5pt}
    \subfigure[Latency in WAN for $n=31$.]{ 
		\input{figs/MIB-latency_n=31}
		\label{fig:MIB-latency(n=31)}
	}
	\hspace{0.5pt}
	\subfigure[Latency in WAN for $n=46$.]{ 
		\input{figs/MIB-latency_n=46}
		\label{fig:MIB-latency(n=46)}
	}
	\hspace{0.5pt}
	\subfigure[Latency in LAN for $f=1$.]{ 
		\input{figs/MIB-latency-lan_f=1}
		\label{fig:MIB-latency-lan(f=1)}
	}
	\hspace{0.5pt}
	\subfigure[Latency in WAN for $f=1$.]{
		\input{figs/MIB-latency-wan_f=1}
		\label{fig:MIB-latency}
	}
	\hspace{0.5pt}
	\subfigure[Latency in WAN for $f=5$]{ 
		\input{figs/MIB-latency_f=5}
		\label{fig:MIB-latency(f=5)}
	}
	\hspace{0.5pt}
	\subfigure[Latency in WAN for $f=10$]{ 
		\input{figs/MIB-latency_f=10}
		\label{fig:MIB-latency(f=10)}
	}
	\hspace{0.5pt}
	\subfigure[Latency in WAN for $f=15$]{ 
		\input{figs/MIB-latency_f=15}
		\label{fig:MIB-latency(f=15)}
	}
	\caption{Latency of \mib protocols and BEAT in both LAN and WAN settings. }
	\label{fig:latency}
\end{figure*}

\heading{Evaluation overview.}
We deploy the \mib protocols and BEAT on Amazon EC2 and evaluate their performance utilizing up to 140 instances distributed evenly in five continents. By default, we use the \textit{t2.medium} type instances. Each instance has two virtual CPUs and 4GB memory. For one set of experiments, we also evaluate the peak throughput using \textit{t2.micro} instances, each of which has one virtual CPUs and 1GB memory. 
The size of each transaction is 250 bytes. In every epoch, each replica proposes $b = \lceil B/n \rceil$  transactions, where $B$ is the batch size. 

We distinguish two scenarios: \textit{the same $n$ setting} and \textit{the same $f$ setting}. The same $n$ setting allows evaluating the performance of all protocols with the same total number of replicas. In a system with $n$ replicas, BEAT tolerates $\lfloor \frac{n-1}{3} \rfloor$ failures, \miba and its variants tolerate $\lfloor \frac{n-1}{5} \rfloor$ failures, and \mibb and its variants tolerate $\lfloor  \frac{n-1}{7} \rfloor$ failures. In our evaluation, we choose $n = 16$, $31$, and $46$. For instance, when $n =31$, BEAT tolerates 10 failures, \miba and its variants tolerate 6 failures, and \mibb and its variants tolerate 4 failures. The same $n$ setting can provide guidance for selecting protocols when one has a fixed number of nodes for an application. For instance, if one has 31 nodes and is certain there would not be 6 failures, he or she may favor \miba over BEAT for performance considerations.

The same $f$ setting enables us to assess the performance for systems tolerating the same number of failures $f$. In this setting, the total number of replicas for BEAT, \miba and its variants, and \mibb and its variants is $3f+1$, $5f+1$, and $7f+1$, respectively. The evaluation for the same $f$ setting is important for three reasons.  
First, when one designs systems with a particular goal of tolerating some $f$ failures, the evaluation can be directly used as a guideline. In many cases, one may not wish to adopt a system with fewer replicas, as other systems with more replicas may be more efficient. 
Second,  it helps understand the performance difference among \miba and its variants, as well as the difference among \mibb and its variants. Indeed, such an evaluation allows comparing protocols with different RBC and ABA components, thereby validating our theoretical design. 
Third, it enables us to analyze some corner cases: for instance, if one has 6 nodes, the system can only tolerate $f = 1$ failure, whether using BEAT or \miba. The evaluation for the $f=1$ case can help users to understand the trade-offs.   

For some experiments, we vary the size of $b$ from 1 (250 bytes per replica) to 10000 (2.38 MB per replica) to evaluate the throughput. We evaluate the latency when there is no contention, i.e., when $b=1$. We evaluate the performance in the LAN settings (where the nodes are launched in the same EC2 region) and the WAN settings (where the nodes are evenly distributed in five continents).

\begin{figure*}[ht!]
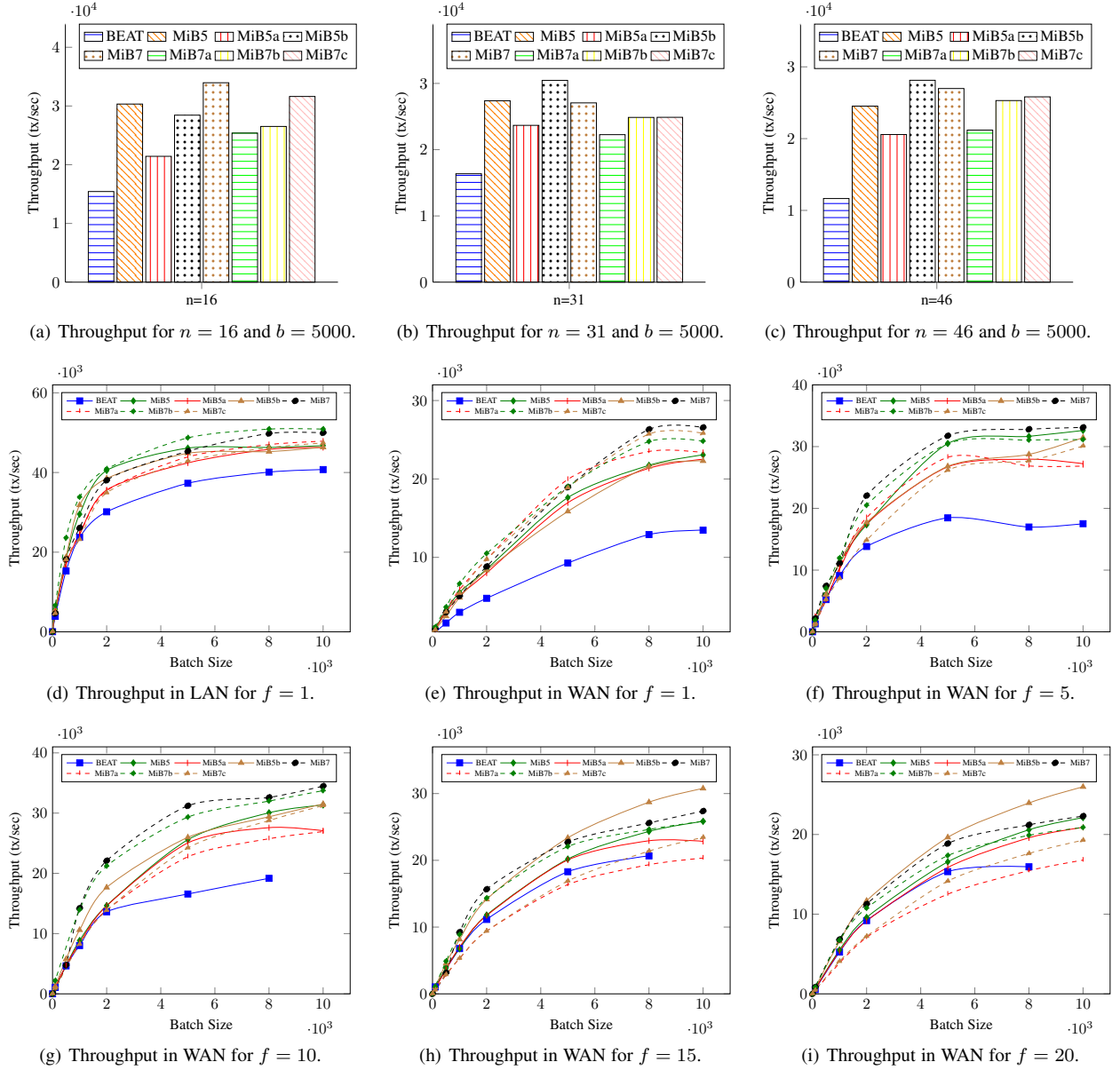

\centering
     \subfigure[Throughput for $n=16$ and $b=5000$.]{
		\input{figs/MIB-Throughput_n=16}
		\label{fig:MIB-Throughput(n=16)}
	}
	\hspace{1pt}
	 \subfigure[Throughput for $n=31$ and $b=5000$.]{
		\input{figs/MIB-Throughput_n=31}
		\label{fig:MIB-Throughput(n=31)}
	}
	\hspace{1pt}
	 \subfigure[Throughput for $n=46$ and $b=5000$.]{
		\input{figs/MIB-Throughput_n=46}
		\label{fig:MIB-Throughput(n=46)}
	}
	\hspace{1pt}
	\subfigure[Throughput in LAN for $f = 1$.]{
		\input{figs/MIB-throughput-lan_f=1}
		\label{fig:MIB-throughput-lan(f=1)}
	}
	\hspace{1pt}
	\subfigure[Throughput in WAN for $f = 1$.]{
		\input{figs/MIB-throughput-wan_f=1}
		\label{fig:MIB-throughput-wan(f=1)}
	}
	\hspace{1pt}
	\subfigure[Throughput in WAN for $f = 5$.]{
		\input{figs/MIB-throughput-wan_f=5}
		\label{fig:MIB-throughput-wan(f=5)}
	}
	\hspace{1pt}
	\subfigure[Throughput in WAN for $f = 10$.]{
		\input{figs/MIB-throughput-wan_f=10}
		\label{fig:MIB-throughput-wan(f=10)}
	}
	\hspace{1pt}
	\subfigure[Throughput in WAN for $f = 15$.]{
		\input{figs/MIB-throughput-wan_f=15}
		\label{fig:MIB-throughput-wan(f=15)}
	}
	\hspace{1pt}
	\subfigure[Throughput in WAN for $f = 20$.]{
		\input{figs/MIB-throughput-wan_f=20}
		\label{fig:MIB-throughput-wan(f=20)}
	}
	\caption{Throughput of \mib protocols and BEAT in the LAN and WAN settings.}
	\label{fig:throughputsamen}
\end{figure*}

\subsection{Latency}
\heading{The same $\emph{n}$}. We first compare the latency of the protocols in the WAN setting when $n$ is fixed. As shown in Figure 10(a-c), all \mib protocols have significantly lower latency than BEAT. This is expected, since all \mib protocols terminate in fewer steps than BEAT (for both the best-case and average-case scenarios). For instance, when $n=16$, the latency of \miba is $63.5\%$ of that in BEAT. When we have a larger $n$, the difference between BEAT and \miba is more visible. For $n=46$, the latency of \miba is $69.1\%$ of that in BEAT. 

\mibb and its variants have consistently higher latency than \miba and its variants. When $n=16$, the latency of \mibb is $25.9\%$ higher than \miba. This is mainly because replicas need to collect more matching messages in both RBC and ABA phases. \mibb protocols need to collect increasingly more messages when $n$ grows larger and the latency difference between the \mibb instance and \miba is more significant.

We find that the two \miba variants have consistently higher latency than \miba.
This does not hold for the \mibb variants. Interestingly, we observe that the result depends on the size of $b$. When $b$ is small, both \mibba and \mibbb have lower latency than \mibb and \mibbc. Indeed, both \mibb and \mibbc are RBC with learners involving an additional step for some replicas. When $b$ grows larger, the network bandwidth consumption dominates the overhead, and correspondingly, \mibb and \mibbc have lower latency. 

\heading{The same $\emph{f}$}. Figure 10(d-h) show the latency of the protocols in the LAN and WAN settings for the same~$f$. Due to the upper bound on the number of EC2 instances we can launch in an EC2 region, we can evaluate the latency in the LAN setting only when $f=1$. As shown in \figref{fig:MIB-latency-lan(f=1)}, not surprisingly, the latency for all protocols is lower than the results in the WAN environment. BEAT has lower latency than other \mib protocols. Indeed, all \mib protocols have more replicas given the same $f$. For instance, when $f=1$, \mibb has 8 replicas, while BEAT has 4 replicas. Replicas in \mibb and BEAT need to collect 7 and 3 matching messages in each RBC and ABA invocation, respectively. Thus, \mibb and its variants have higher latency. The same result applies to the results in the WAN environment and when $f$ grows larger. Specifically, all \mibb protocols have higher latency than \miba and BEAT. \miba and its variants have higher latency than BEAT, except that \mibaa has lower latency than BEAT for the $f=1$ case only.

\begin{figure*}[ht]
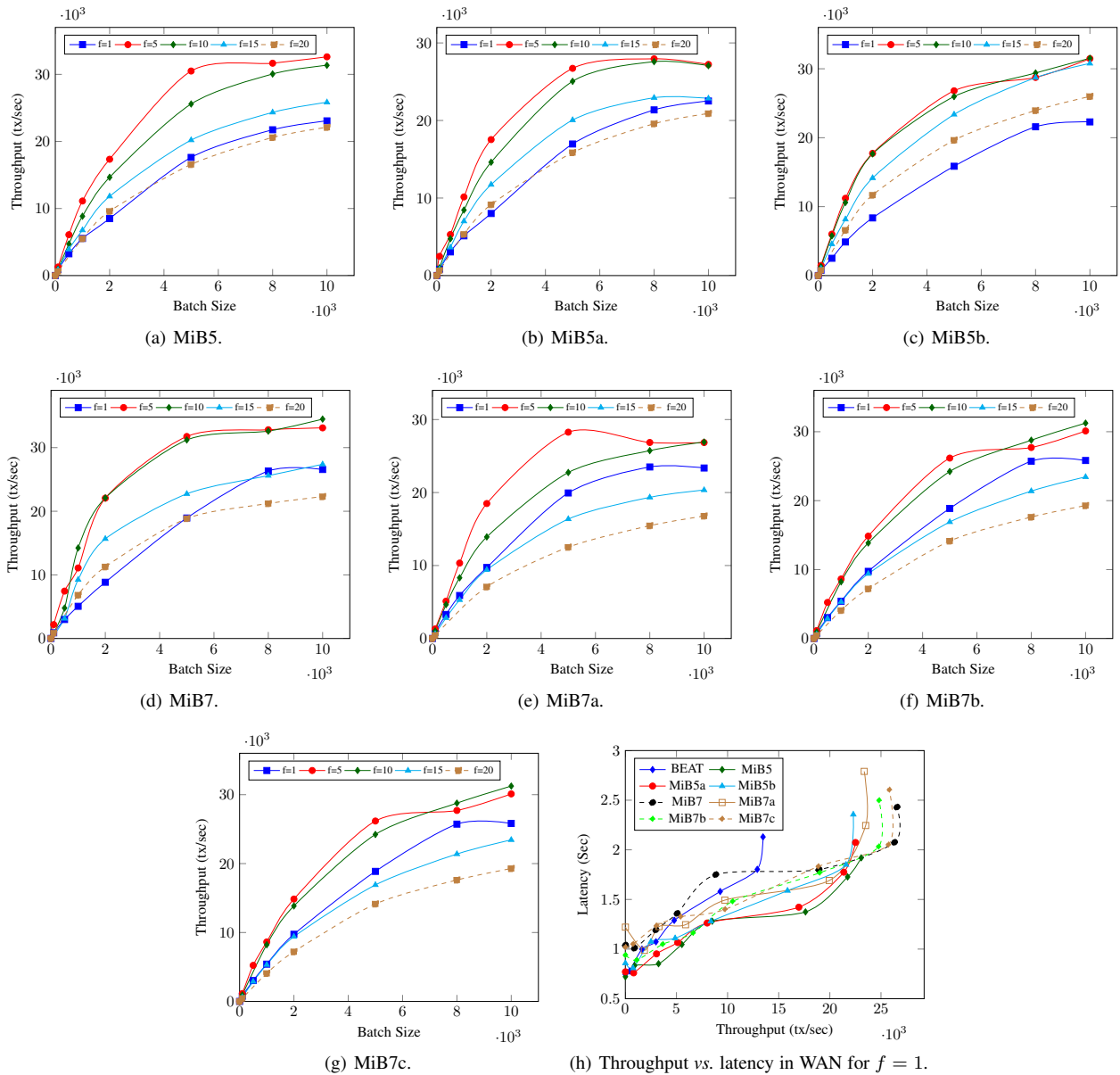

	\centering
	\hspace{1pt}
	\subfigure[\miba.]{
		\input{figs/MIB-scalability-MIB2}
		\label{fig:MIB2}
	}
	\hspace{1pt}
	\subfigure[\mibaa.]{
		\input{figs/MIB-scalability-MIB1}
		\label{fig:MIB1}
	}
	\hspace{1pt}
	\subfigure[\mibab.]{
		\input{figs/MIB-scalability-MIB3}
		\label{fig:MIB3}
	}
	\hspace{1pt}
	\subfigure[\mibb.]{
		\input{figs/MIB-scalability-MIB7}
		\label{fig:MIB7}
	}
	\hspace{1pt}
	\subfigure[\mibba.]{
		\input{figs/MIB-scalability-MIB4}
		\label{fig:MIB4}
	}
	\hspace{1pt}
	\subfigure[\mibbb.]{
		\input{figs/MIB-scalability-MIB6}
		\label{fig:MIB5}
	}
		\hspace{1pt}
	\subfigure[\mibbc.]{
		\input{figs/MIB-scalability-MIB6}
		\label{fig:MIB6}
	}
	\hspace{1pt}
	\subfigure[Throughput \textit{vs.} latency in WAN for $f = 1$. ]{
		\input{figs/MIB-latencyvsthroughput-Lan_f=1}
		\label{fig:MIB-latencyvsthroughput-Lan(f=1)}
	}
	\caption{(a-g) Scalability of \mib protocols; (h) throughput vs. latency in WAN.}
    \label{fig:scalability}
	\vspace{-1.8em}
\end{figure*}

\subsection{Throughput}

\noskipheading{The same $\emph{n}$}. We assess the throughput of the protocols using the same $n$. We fix the batch size to 5000 and evaluate the throughput. As shown in Figure 11(a-c), all \mib protocols achieve significantly higher throughput than BEAT. As an example, when $n=16$, the throughput of \miba is $96.5\%$ higher than that of BEAT, while the throughput of \mibb is $120.0\%$ higher than that of BEAT. When $n=46$, \miba and \mibb achieve $110.4\%$ and $131.5\%$ higher throughput than BEAT, respectively. The performance improvement is mainly due to the step reduction in both the RBC and ABA components.

\heading{The same $\emph{f}$}. We report the throughput of the protocols with the same $f$ from Figure 11(d-i). In both the LAN and WAN settings, when $f$ is no greater than 10, all \mib protocols achieve consistently higher throughput than BEAT. Even when $f$ is greater than $10$, BEAT outperforms \mibba and \mibbc only in some cases; other \mib protocols are consistently more efficient than BEAT. The performance difference is, in part, because \mib reduces the number of steps. Furthermore, given the same $f$, \mib protocols have a larger $n$ and can propose more concurrent transactions. For instance, when $f=1$ and $b=5000$, the number of proposed transactions for BEAT is $33.3\%$ and $50\%$ less than \miba and \mibb, respectively. We also find that when $f$ grows larger (when $f \geq 15$), the network bandwidth consumption dominates the overhead and correspondingly the performance difference between BEAT and \mib protocols becomes comparatively small. 
\begin{figure*}[b!]
\centering
    
	\subfigure[Throughput in LAN for $f = 1$ in failure-free scenario.]{
		\input{figs/f=1Lannofailure}
		\label{fig:MIB-throughput-f=1Lannofailure}
	}
	\hspace{1pt}
	\subfigure[Throughput in LAN for $f = 1$ in crash failure scenario.]{
		\input{figs/f=1Lanstop}
		\label{fig:MIB-throughput-f=1Lanstop}
	}
	\hspace{1pt}
	\subfigure[Throughput in LAN for $f = 1$ in Byzantine scenario.]{
		\input{figs/f=1Lanfail}
		\label{fig:MIB-throughput-f=1Lanfail}
	}
	\hspace{1pt}
	\subfigure[Throughput in WAN for $f = 1$ in failure-free scenario.]{
		\input{figs/f=1Wannofailure}
		\label{fig:MIB-throughput-f=1Wannofailure}
	}
	\hspace{1pt}
	\subfigure[Throughput in WAN for $f = 1$ in crash failure scenario.]{
		\input{figs/f=1Wanstop}
		\label{fig:MIB-throughput-f=1Wanstop}
	}
	\hspace{1pt}
	\subfigure[Throughput in WAN for $f = 1$ in Byzantine scenario.]{
		\input{figs/f=1Wanfail}
		\label{fig:MIB-throughput-f=1Wanfail}
	}
	\hspace{1pt}
	\subfigure[Throughput in WAN for $f = 5$ in failure-free scenario.]{
		\input{figs/f=5Wannofailure}
		\label{fig:MIB-throughput-f=5Wannofailure}
	}
	\hspace{1pt}
	\subfigure[Throughput in WAN for $f = 5$ in crash failure scenario.]{
		\input{figs/f=5Wanstop}
		\label{fig:MIB-throughput-f=5Wanstop}
	}
	\hspace{1pt}
	\subfigure[Throughput in WAN for $f = 5$ in Byzantine scenario.]{
		\input{figs/f=5Wanfail}
		\label{fig:MIB-throughput-f=5Wanfail}
	}
	\hspace{1pt}
	\subfigure[Throughput in WAN for $n=16$ in failure-free scenario.]{
		\input{figs/n=16Wannofailure}
		\label{fig:MIB-throughput-n=16Wannofailure}
	}
	\hspace{1pt}
	\subfigure[Throughput in WAN for $n=16$ in crash failure scenario.]{
		\input{figs/n=16Wanstop}
		\label{fig:MIB-throughput-n=16Wanstop)}
	}
	\hspace{1pt}
	\subfigure[Throughput in WAN for $n=16$ in Byzantine scenario.]{
		\input{figs/n=16Wanfail}
		\label{fig:MIB-throughput-n=16Wanfail}
	}

	\label{fig:performanceunderfailures}
\end{figure*}

We also report in \figref{fig:MIB-latencyvsthroughput-Lan(f=1)} throughput vs. latency in the WAN setting for $f=1$. 

\heading{\miba \textit{vs.} \mibb}. We first report the throughput for the same $n$. When $n=16$, \mibb outperforms other protocols. When $n=31$ and $n=46$, \mibab outperforms all other \mib protocols. This is expected, since both \mibab and \mibb use learners to reduce the number of messages transmitted. Due to the use of learners, \mibbc also achieves higher throughput among the \mib variants. The reason why \mibbc achieves lower throughput than \mibab and \mibb is that \mavid has one more step than \mbc. 

For all experiments, \mibaa and \mibba achieve lower throughput than the other \mib protocols. Note \mibaa and \mibba use \mavid and do not use learners; \mavid has one more step and more messages than \mbc. 

We also evaluate the performance using the same $f$. When $f=1$ in the LAN environment, \mibbb outperforms other protocols. This is mainly because more transactions are proposed with a larger $n$. In contrast, in the WAN setting, \mibb outperforms other protocols when $f \leq 10$ and \mibab outperforms other protocols when $f \geq 15$. In particular, both \mibab and \mibb use learners to reduce the number of messages transmitted. Furthermore, \mibb variants in general achieve higher throughput than \miba variants when $f$ is smaller than $10$ and lower throughput than \miba variants when $f$ is greater than $10$.

\heading{\mib with learners}. \mibab, \mibb, and \mibbc use learners and thus outperform other protocols (except for the $f=1$ case). 
 For instance, when $f=1$ in the LAN setting, the peak throughput of \mibbb is $1.84\%$ higher than \mibb. When $f$ equals $5$, $10$, $15$, and $20$ in the WAN setting, the peak throughput of \mibb is $6.3\%$, $2.2\%$, $5.9\%$, and $6.8\%$ higher than \mibbb, respectively.

\subsection{Scalability}
We evaluate the scalability of all our implemented protocols by varying $f$ from 1 to 20 and varying $b$ from 1 to 10000. When $f$ increases from 1 to 20, the throughput first increases and then decreases. As illustrated in \figref{fig:scalability}, \mib protocols achieve their peak throughput around $f=5$ and $f=10$. 
The trend is very similar to that of BEAT (and EPIC): when $f$ increases, the number of proposed transactions also increases so the throughput becomes higher; when $f$ further increases, the network bandwidth becomes the performance bottleneck.

\begin{figure*}[ht!]
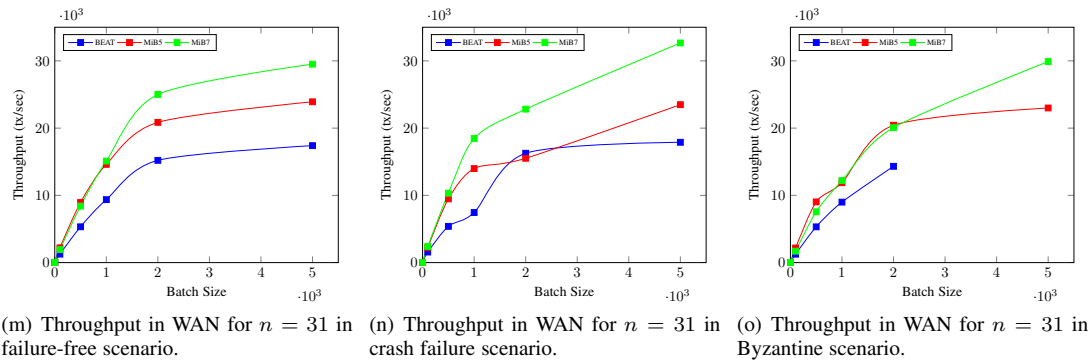

\centering
	\subfigure[Throughput in WAN for $n=31$ in failure-free scenario.]{
		\input{figs/n=31Wannofailure}
		\label{fig:MIB-throughput-n=31Wannofailure}
	}
	\hspace{1pt}
	\subfigure[Throughput in WAN for $n=31$ in crash failure scenario.]{
		\input{figs/n=31Wanstop}
		\label{fig:MIB-throughput-n=31Wanstop}
	}
	\hspace{1pt}
	\subfigure[Throughput in WAN for $n=31$ in Byzantine scenario.]{
		\input{figs/n=31Wanfail}
		\label{fig:MIB-throughput-=31Wanfail}
	}

	\caption{Throughput of BEAT, \miba, and \mibb in the LAN and WAN settings in failure-free, crash failure, and Byzantine scenarios.}
	\label{fig:throughputsamen}
\end{figure*}

\begin{figure*}
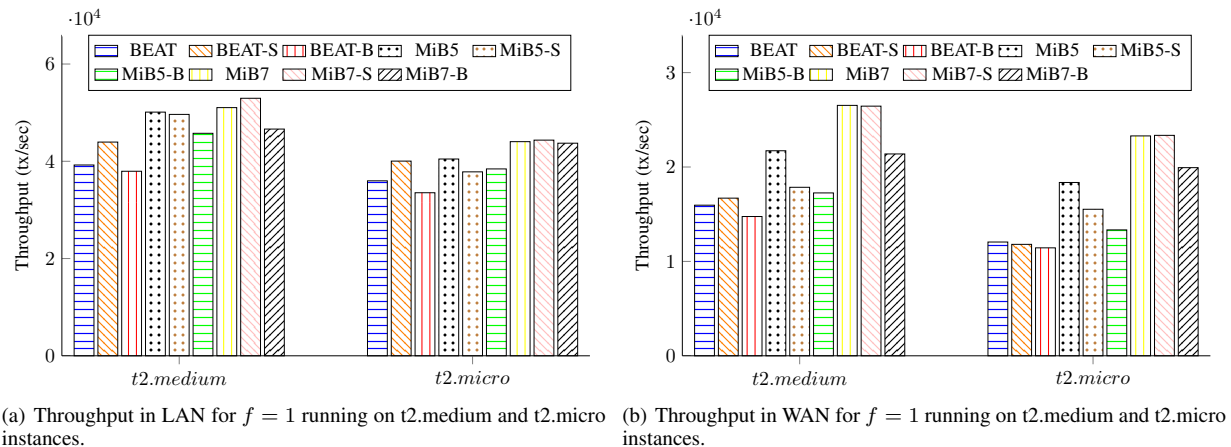

\centering
	\subfigure[Throughput in LAN for $f=1$ running on t2.medium and t2.micro instances.]{
		\input{figs/t2.microLanf=1}
		\label{fig:MIB-throughput-t2.microLanf=1}
	}
	\hspace{1pt}
	\subfigure[Throughput in WAN for $f=1$ running on t2.medium and  t2.micro instances.]{
		\input{figs/t2.microWanf=1}
		\label{fig:MIB-throughput-t2.microWanf=1}
	}
	\caption{Throughput of BEAT, \miba, and \mibb running on different hardware.}
	\label{fig:throughputhardware}
\end{figure*}

\subsection{Performance under Failures}
We now evaluate the performance of the protocols under failures. In \figref{fig:throughputsamen}, we show the throughput of BEAT, \miba and \mibb in three different scenarios: failure-free, crash failure, and Byzantine. In the crash failure scenario, we stop $f$ replicas and run the protocols. In the Byzantine scenario, we chose to \textit{simulate} the Byzantine behavior in the ABA phase instead of the RBC phase, because asynchronous RBC protocols are incredibly robust against Byzantine failures (see~\cite{astro}). For the ABA phase, a Byzantine replica may exhibit either of the two following behaviors: not sending ABA proposals, or sending inconsistent proposals to different replicas. The former has been captured by the crash failure scenario. Hence, we focus on the latter one for Byzantine scenarios, where we let Byzantine replicas propose inconsistent values in the ABA phase. Note that inconsistent votes from Byzantine replicas may cause the protocols to terminate using more rounds, a strategy that might impact performance. Meanwhile, in this scenario, the ABA phase in both \miba and \mibb would switch to a slower protocol, which may further reduce performance.  
We vary $f$ from 1 to 10 in WAN to evaluate the performance. For $f=1$, we also evaluate the throughput in LAN.

\heading{Failure-free vs. crash failure.} All the three protocols achieve higher throughput in the crash failure scenario than that in the failure-free scenario. For instance, the throughput of BEAT in crash failure scenario is 4.6\%-24.5\% higher than that in failure-free scenario. The throughput of \miba in crash failure scenario is 0.3\%-9.4\% than that in the failure-free scenario. For \mibb, the throughput is 1.5\%-10.6\% higher in the crash failure scenario than the failure-free scenario. The results are expected, since the network bandwidth consumption is lower when $f$ replicas crash compared to the failure-free scenario. Compared to BEAT, the throughput improvement under crash failures for both \miba and \mibb are lower. This is because both \miba and \mibb achieve one-step termination in the failure-free cases for over $n-f$ ABA instances. In contrast, in the crash failure scenario, both protocols switch to a slower backup ABA protocol under failures. 

\heading{Failure-free vs. Byzantine.} For all the experiments, all the three protocols achieve higher throughput in the failure-free scenario than the Byzantine scenario. The throughput of BEAT in failure-free scenario is 1.9\%-4.9\% higher than that in the Byzantine scenario. For \miba, the throughput is 2.9\%-16.1\% higher in the failure-free scenario. The throughput of \mibb is 2.2\%-6.0\% higher in the failure-free scenario. The results are also expected since faulty replicas broadcast inconsistent messages to all replicas so the network bandwidth consumption is higher than that in the failure-free scenario. The degradation of performance under Byzantine failures for \miba and \mibb is higher compared to that in BEAT. This is because both protocols switch to a backup ABA under failures.  

\heading{Performance using different hardware}. We assess the performance of the protocols using both \textit{t2.medium} instances and \textit{t2.micro} instances. We show the peak throughput of BEAT, \miba, and \mibb. For each protocol, we evaluate the peak throughput under failure-free, crash failure, and Byzantine scenarios. For each protocol, we use protocol name to represent the performance in failure-free scenario, -S to represent the performance of the crash failure scenario, and -B to represent the performance of the Byzantine scenario. We show the performance in \figref{fig:throughputhardware}.

All the protocols achieve higher throughput using \textit{t2.medium} instances. This is expected since all the computational operations are faster. In the LAN setting, the performance improvement BEAT in all three scenarios are consistency lower than that in the WAN setting. For instance, BEAT, BEAT-S, and BEAT-B achieve 9.0\%, 9.8\%, and 13.1\% higher throughput using \textit{t2.medium} than that in \textit{t2.micro} in LAN, separately. In the WAN setting, the throughput are 32.3\%, 41.5\%, and 29.0\% higher in the three scenarios. In contrast, the performance improvement for both \miba and \mibb are in general higher in the WAN setting. For instance, \miba achieves 23.8\% and 18.3\% higher throughput using \textit{t2.medium} in LAN and WAN, separately. \mibb achieves 15.9\% and 13.2\% higher throughput using \textit{t2.medium} in LAN and WAN, separately. This is mainly because both \miba and \mibb involve more replicas ($5f+1$ and $7f+1$) so the network bandwidth consumption dominates the overhead of the protocols.

\section{Conclusion}\label{conclusion}

We study two important directions in asynchronous BFT---BFT with suboptimal resilience and BFT performance under failures and attacks. This paper provides \mib, a novel and efficient asynchronous BFT framework using new distributed system constructions as building blocks (including an erasure-coded version of IR RBC and a learner-version of RBC). \mib consists of two main BFT instances and five other variants. We design experiments with failures and systematically evaluate the performance of asynchronous BFT protocols (including MiB) in crash failure and Byzantine failure scenarios. Via a five-continent deployment using 140 replicas, we show the \mib instances have lower latency and much higher throughput than their asynchronous BFT counterparts, and moreover, asynchronous BFT protocols, including our \mib protocols, are indeed robust against failures and attacks.

\bibliographystyle{abbrv}
\bibliography{epic}

\vspace{-1cm}

\end{document}